# Ultra-directional high-efficient chiral silicon photonic circuits

Liang Fang[1], Hao-Zhi Luo[2], Xiao-Ping Cao[1], Shuang Zheng[1], Xin-Lun Cai[2,3], and Jian Wang[1,*]

[1]*Wuhan National Laboratory for Optoelectronics, School of Optical and Electronic Information, Huazhong University of Science and Technology, Wuhan 430074, Hubei, China.*
[2]*State Key Laboratory of Optoelectronic Materials and Technologies and School of Physics and Engineering, Sun Yat-sen University, Guangzhou 510275, China*
[3]*caixlun5@mail.sysu.edu.cn*
**Corresponding author: jwang@hust.edu.cn*

**Chiral light-matter interaction enables new fundamental researches and applications of light. The interaction has traditionally faced challenges in low directionality and efficiency based on spin-orbit interaction of light in microscopic waveguides. It is pivotal to exploit photonic integrated circuits to efficiently engineer photonic chiral behavior. Here, we present ultra-directional high-efficient chiral coupling in silicon photonic circuits based on low-order to high-order mode conversion and interference. We show that the directionality of chiral coupling, in principle, can approach ±1 with circular polarization inputs, benefited from the underlying mechanism of complete destructive and constructive interference. The chiral coupling efficiency can exceed 70%, with negligible scattering to non-guided modes, much higher than conventional coupling mechanisms. Moreover, the chiral silicon photonic circuits can function as a perfect 3-dB power splitter for arbitrarily linear polarization inputs, and also open up the possibility of on-chip chirality determination to further flourish the development of chiral optics.**

## 1. INTRODUCTION

Chiral optics is currently flourishing in the fundamental researches and new applications of light. It is associated with optical polarization handedness or spin angular momentum (SAM) that determines the chiral (spin-dependent) behavior of light-matter interactions. Analogous to electric spin Hall effect (SHE) characterizing spin-dependent transport of electrons, as well as quantum SHE version related to unidirectional edge spin transport, the new-found photonic counterparts called the photonic (quantum) SHEs feature the representative phenomenon of chiral optics [1-3]. It emerges as the manifestation of chiral splitting or shift of light beams and spin-controlled unidirectional excitation of surface plasmon-polariton or waveguide modes. The physical mechanism behind them is the general spin-orbit interaction of light in metasurfaces, gradient-index media or strongly confined nano-waveguides [4-9]. In particular, the chiral effects of light-matter interactions in photonic nanostructures may offer a robust opportunity for developing chiral quantum optics and thus promoting the quantum information processing and computing [10-15]. Additionally, the potential applications and researches related with chiral optics cover chiral imaging [16], optical storage [17], all-optical magnetic recording [18], and valley information processing [19,20].

The photonic chiral behavior has been immensely studied on the platforms of metasurfaces [21], nanostructures [7,9], and various optical interfaces, such as air-glass [22] and metal-dielectric interfaces [5]. As for the nanophotonic waveguides, the strongly confined guided-modes naturally manifest as nonnegligible longitudinal polarization component. Accordingly, it gives rise to the large intrinsically transverse spin, and thus induces the remarkable spin-momentum locking phenomenon of light [23-25]. Based on this effect of nanophotonic waveguides, the on-chip chiral resolving, chiral photonic circuit emission and non-reciprocal phenomenon have been recently revealed and investigated via silicon microdisk [26], dipole emission [14], and photonic crystal waveguides with embedded quantum dots [11,15]. It is worthy of note that most of the reported photonic chiral behaviors face great challenges in low directionality and low efficiency, fundamentally limited by their chiral mechanisms. Because there exits substantial coupling to non-guided modes, it shows low chiral coupling efficiency by dipole emitter and silicon microdisk scatter [5,7,26], except for chiral emission utilizing quantum dots, but with requirement of inducing magnetic field [11]. As well known, for the scalable technology of photonic integration and even the future quantum internet, silicon photonics provides a promising platform to solve the problems of miniaturization, fabrication cost, and compatibility with mature complementary metal-oxide-semiconductor (CMOS) technologies [27,28]. Over the past few decades, the conventional fundamental transverse-electrical (TE) and transverse-magnetic (TM) mode management in silicon photonic circuits has been well studied, such as linear polarization (LP) beam splitting/rotating [29] and sorting [30]. However, the analogue for managing circularly polarized light has received little attention on silicon platforms. Moreover, return to chiral optics scheme, it is also significant to exploit a new method to engineer photonic chiral behavior in silicon photonic circuits.

Here we demonstrate an ultra-directional and high-efficient chiral coupling in silicon photonic circuits based on low-order to high-order mode conversion and interference. The chiral coupling in photonic integrated circuits manifests as remarkably directional coupling that completely depends upon the polarization handedness of the incident light. It removes the conventional coupling restriction with only LP inputs in silicon photonic circuits. The underlying mechanism enabling chiral coupling is optical interference that is a well-known phenomenon and the cornerstone of numerous applications in optics. The spin-controlled directional coupling based on interference principle has previously been demonstrated for plasmon polaritons emission [5,31]. We exploit the interference with different mode orders to achieve high directionality of chiral coupling because of the complete destructive and constructive interference. Furthermore, the chiral coupling based on the guided-mode interference possesses high efficiency, notably much higher than the mechanism originated from the spin-momentum locking of light at the nanowaveguide interfaces. Moreover, the proposed chiral silicon photonic circuits enable to on-chip exact chirality determination for polarization handedness of light, and could also function as a perfect 3-dB power splitter for LP incident light with arbitrary polarization orientation, which has not yet been demonstrated before in the polarization-sensitive silicon nanophotonic devices.

## 2. PRINCIPLE AND DESIGN

Firstly, we retrospect the study of helicity of optical polarization handedness. The complex electric-field of light with polarization handedness can be described as

$$\mathbf{E} = A\left[\vec{e}_x + m \cdot \exp(-i\delta) \cdot \vec{e}_y\right]\exp\left[i(kz - \omega t)\right], \quad (1)$$

where $A$ is the wave amplitude, $\vec{e}_x$ and $\vec{e}_y$ denote the unit vectors of the corresponding axes, $m$ indicates the amplitude ratio of y-polarized and x-polarized electric-field components, $\delta$ describes the phase retardation between them, $k = \omega/c$ is the wave number with $\omega$ being angular frequency and $c$ being the speed of light in vacuum. The helicity of polarization handedness is given by [23,32]

$$\sigma = \frac{2 \cdot \mathrm{Im}\left(E_x \cdot E_y^*\right)}{\left|E_x\right|^2 + \left|E_y\right|^2} = \frac{2m \cdot \sin\delta}{1 + m^2}. \quad (2)$$

Here $\sigma \in [-1, 1]$, especially, the integers $\sigma = -1, 1$ and 0 describe the left-handed circular polarization (LCP) $\sigma^{-1}$, right-handed circular polarization (RCP) $\sigma^{+1}$, and LP states of light, respectively, correspondingly carrying mean SAM of $-\hbar$, $+\hbar$ and 0 per photon, where $\hbar$ denotes Planck's constant divided by $2\pi$.

The chiral effect of light-matter interactions is generally illustrated in Fig. 1(a). It is characterized by spin-dependent splitting of light or directional emitting, scattering and coupling of light. Its mechanisms include wave interference [5,31], spin-orbit coupling [4,8], and spin-momentum locking [2,7,11] of light. In our scheme, the chiral photonic device is formed on a silicon-on-insulator (SOI) platform. As shown in Fig. 1(b), the incident light with LCP or RCP state is injected into a polymer (SU8)-assisted inversely tapered Y-branch silicon waveguide for chiral coupling. This specific waveguide structure for chiral coupling can be divided into three parts, i.e. a thick wire polymer waveguide covering an inversely tapered silicon waveguide at the bottom (part I), a subsequent adiabatic inverse taper structure after the polymer waveguide (part II), and a Y-branch waveguide at the end of the inverse taper structure (part III). The insets in Fig. 1(b) show zoom-in details of the three parts of the designed device.

The working principle of the chiral silicon photonic device relies on low-order to high-order mode conversion and interference, briefly described as follows. Despite the mirror symmetry relative to the YZ plane, the waveguide structure, consisting of upper-cladding of air and buffer layer of $SiO_2$, breaks the mirror symmetry relative to the XZ plane [30]. Such asymmetry could be understood as a fundamental requirement for chirality sorting with circular polarization. For the incident light, the x-polarization component excites the fundamental $TE_0$ mode of the polymer waveguide with high efficiency. It is then coupled into the inversely tapered silicon waveguide sitting at the bottom of the polymer waveguide, and maintained in a subsequent adiabatically tapered silicon waveguide as $TE_0$ mode. Meanwhile, the y-polarization component excites the $TM_0$ mode coupling from the polymer waveguide to the inversely tapered silicon waveguide at the bottom. The subsequent adiabatic inverse taper structure after the polymer waveguide converts the $TM_0$ mode into the first-order $TE_1$ mode because of the mode hybridization with structure asymmetry relative to the XZ plane [33], as shown in Fig. 1(c), which enables the transfer of guided modes from the quasi-vertical polarization to the quasi-horizontal polarization.

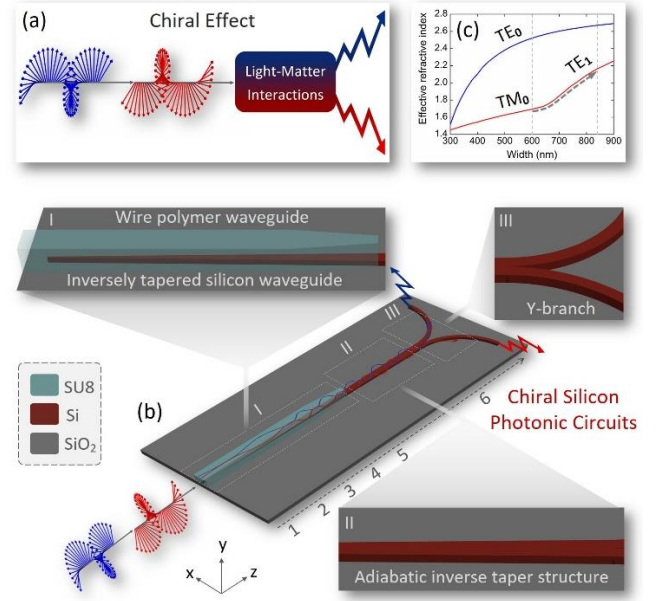

**Fig. 1.** Principle and design of chiral silicon photonic circuits. (a) Schematic illustration of chiral effect of light-matter interactions. The polarization handedness of incident light determines the chiral behavior of light-matter interactions. (b) 3D view of chiral silicon photonic circuits (polymer-assisted inversely tapered Y-branch silicon waveguide) for chiral coupling based on low-order to high-order mode conversion and interference. (c) Calculated effective refractive index versus the silicon waveguide width. In the adiabatic inverse taper structure with the waveguide width varying from 600 to 840 nm, the $TE_0$ mode remains unchanged, while the $TM_0$ mode is evolved into the $TE_1$ mode due to mode hybridization.

The simulated transverse mode pattern evolution at six positions (1, 2, 3, 4, 5, 6 marked in Fig. 1(b)) along the waveguide is manifested in Fig. 2(a) for the incident x-polarized and y-polarized light, respectively. In the adiabatic inverse taper structure, $TE_0$ and $TE_1$ modes both with quasi-horizontal polarization produce spatial interference in the case of circular polarization inputs, giving rise to quasi-periodic up and down oscillation of field density along the propagation direction (Fig. 1(b)). The power evolution of up and down oscillated interference fields can be deduced by making an integration of the field density along the upper and lower half of the field region (see Supplement 1) as follows

$$I_u = \xi + m^2 \cdot \psi + 2m \cdot \zeta \cdot \cos(\Delta\beta \cdot z - \delta), \quad (3)$$

$$I_d = \xi + m^2 \cdot \psi - 2m \cdot \zeta \cdot \cos(\Delta\beta \cdot z - \delta), \quad (4)$$

Here $\xi$ and $\psi$ are integral coefficients for $TE_0$ and $TE_1$ modes on upper and lower half of the field region, respectively, and $\zeta$ indicates that of the superposition term. $\Delta\beta = \Delta n_{eff} k$ is the propagation constant difference between $TE_0$ and $TE_1$ modes at the end of the adiabatic inverse taper structure with $\Delta n_{eff}$ being the effective index difference. Accordingly, the up and down oscillated interference fields completely determine the directional coupling to output of the followed Y-branch waveguide. Note that similar function of Y-branch waveguide might be achieved by multimode interference (MMI) couplers [34,35].

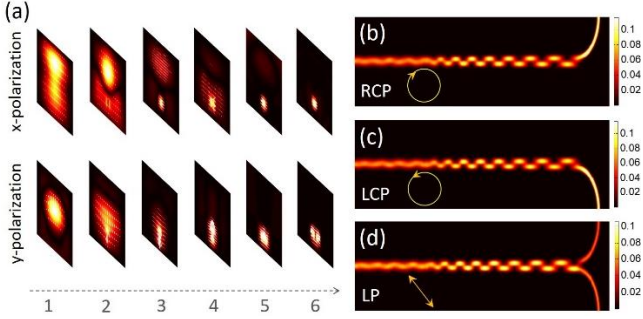

**Fig. 2.** Numerical simulation results. (a) Transverse mode pattern evolution at six different positions (1, 2, 3, 4, 5, 6 in Fig. 1(b)) along the waveguide for x- and y- polarization component inputs. (b)-(d) Simulation results of chiral coupling in silicon photonic circuits when the incident polarization handedness is RCP (b), LCP (c) and LP (d), respectively.

The power splitting here highly depends upon the phase retardation $\delta$ and amplitude ratio $m$, and thus is eventually associated with the helicity ($\sigma$) of the incident light. The directionality of chiral coupling is evaluated by

$$D = \frac{I_u - I_d}{I_u + I_d} = \frac{2m \cdot \zeta \cdot \cos(\Delta\beta \cdot z - \delta)}{\xi + m^2 \cdot \psi}, \quad (5)$$

Remarkably, a high directionality of chiral coupling for $\sigma^{-1}$ and $\sigma^{+1}$ polarization handedness inputs ($m=1$) can be obtained provided that the interference is complete destructive and constructive on the condition of $\xi \simeq \psi \simeq \zeta$ (see Supplement 1). Under such approximation, when suitably setting the joint location of the Y-branch structure to make $\Delta\beta \cdot z = (2n+1/2)\pi$ ($n=0, 1, 2 \ldots$), the directionality can be simplified as,

$$D \simeq \frac{2m \cdot \sin\delta}{1+m^2} = \sigma, \quad (6)$$

which is directly linked to the helicity of polarization handedness of the incident light. Analogously, when $\Delta\beta \cdot z = (2n-1/2)\pi$ ($n=0, 1, 2 \ldots$), the directionality is given by $D \simeq -\sigma$. This is the primary goal of the work. It is noted that the chiral coupling for circular polarization states can be extended to any two orthogonal polarization states such as elliptical polarizations based on the similar design principle (see Supplement 1).

We numerically verify the chiral coupling and show results by three-dimensional finite-difference time domain (3D-FDTD) simulations for three different polarization handedness inputs, as shown in Fig. 2(b)-2(d). For the $\sigma^{+1}$ RCP incident light ($\delta = \pi/2$), $I_u$ is maximal and $I_d$ minimal, hence, $D \simeq 1$ (Fig. 2(b)). For the $\sigma^{-1}$ LCP incident light ($\delta = -\pi/2$), $I_u$ is minimal but $I_d$ maximal, and thus $D \simeq -1$ (Fig. 2(c)). For the LP incident light with $\pm 45°$ diagonal orientation ($\delta = 0$ or $\pi$), $I_u \simeq I_d$, giving $D \simeq 0$ (Fig. 2(d)). Actually, the LP states have other polarization orientations, such as the pure x-polarization ($m=0$) and y-polarization ($m \to \infty$) ones, but all these cases of LP inputs feature similar phenomena as that in Fig. 2(d) with $D \simeq 0$ (see Supplement 1). From another perspective, LP state with orientation angle $\alpha$ can be regarded as the linear combination of orthogonal LCP and RCP. The expansion using Jones vector can be explicitly written as

$$\begin{bmatrix}\cos\alpha \\ \sin\alpha\end{bmatrix} = \frac{1}{2}(\cos\alpha - i\sin\alpha)\begin{bmatrix}1 \\ i\end{bmatrix} + \frac{1}{2}(\cos\alpha + i\sin\alpha)\begin{bmatrix}1 \\ -i\end{bmatrix}, (7)$$

Hence, the chiral silicon photonic device can act as a perfect 3-dB power splitter for arbitrary LP inputs in the case of high-directional chiral coupling for circular polarization inputs. Apparently, this proposed chiral coupling scheme also enables to determinate the polarization handedness of light.

## 3. EXPERIMENTAL RESULTS

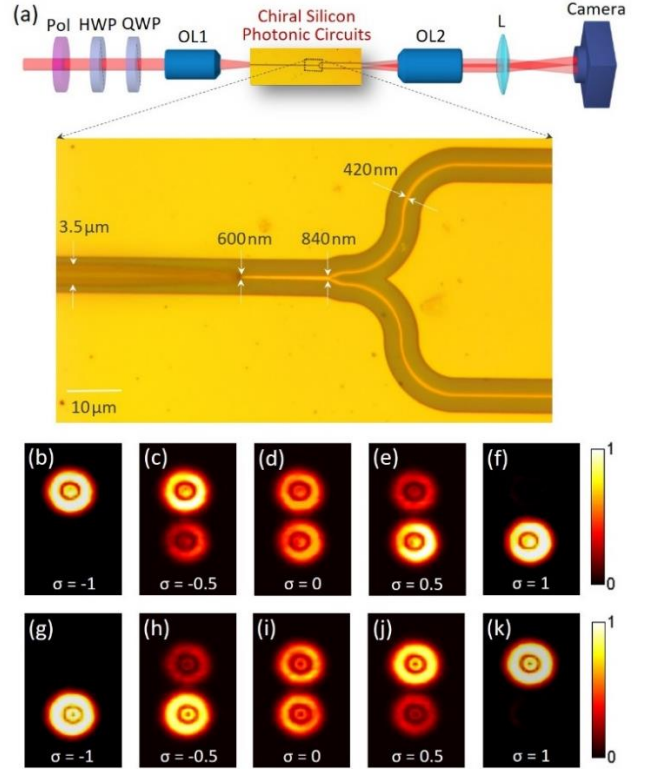

**Fig. 3.** Experimental setup and results to measure the chiral coupling outputs. (a) Experimental schematic diagram with an inset of optical microscope image. Pol: polarizer; HWP: half-wave plate; QWP: quarter-wave plate; OL: objective Lens; L: lens. (b)-(k) Measured spin-dependent output from chiral silicon photonic circuits under different incident polarization handedness with helicity $\sigma$ =-1 (b, g), -0.5 (c, h), 0 (d, i), 0.5 (e, j) and 1 (f, k), respectively, at two wavelengths ($\lambda_1$: (b)-(f), $\lambda_2$: (g)-(k)) with opposite directionality.

We fabricate the chiral photonic device on a silicon platform (see Supplement 1) and demonstrate the chiral coupling in the fabricated silicon photonic circuits. The experimental setup and results are shown in Fig. 3 to measure the chiral coupling outputs from the silicon photonic circuits. The inset of Fig. 3(b) shows the measured optical microscope image of the fabricated chiral silicon photonic circuits (silicon thickness: 220 nm). The input and two output ports of the silicon photonic circuits are all covered by a square polymer (SU8) waveguide (3.5 μm × 3.5 μm) to facilitate efficient excitation and output of light. Note that the polymer waveguide at the two output ports is not shown in Fig. 1(b) for simplicity. The silicon waveguide (length: 240 μm) covered by the input polymer waveguide is inversely tapered with its width slowly increased from 80 nm to 600 nm, enabling high-efficient mode coupling into the waveguide. After exiting from the polymer waveguide, the silicon waveguide is further adiabatically tapered with a length of 16 μm and the varying width from 600 nm to 840 nm, enabling the conversion from $TM_0$ mode to $TE_1$ mode for the y-polarization component of incident light. At the branching point of Y-branch waveguide, two output branches are equally split with small bending radii (bending radius: 1.5 μm), and then connected with the large bending waveguides to guarantee a relatively large distance (~50 μm) between two output branches. Note that the initial bending radii of two output branches need to be small enough to effectively separate the oscillated interference fields and thus achieve high directionality of chiral coupling, despite possibly increasing the insertion loss of the device.

In the measurement setup for free-space coupling, the polarization handedness of incident light is controlled by a group of polarizer and wave plates. The quarter-wave plate (QWP) determines the helicity of polarization handedness as $\sigma \simeq \sin 2\theta$ (see Supplement 1), where $\theta$ is the rotation angle between the optical axis of QWP and the polarization direction of light after half-wave plate (HWP). For light coupling from free-space to waveguide, an objective lens (OL) is used to focus the incident light on the facet of the polymer waveguide, and vice versa, for light output from the waveguide.

The handedness-dependent output from chiral silicon photonic circuits is photographed by a camera under different polarization handedness of incident light with helicity $\sigma$ = -1, -0.5, 0, 0.5 and 1, respectively, at two wavelengths of $\lambda_1$ in Figs. 3(b)-3(f) and $\lambda_2$ in Figs.

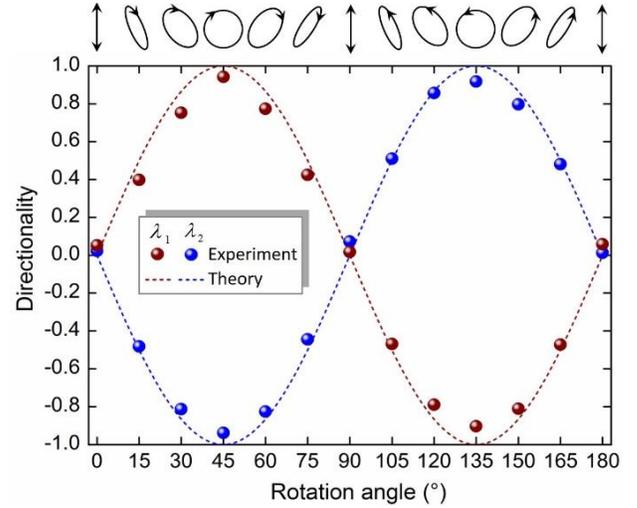

**Fig. 4.** Measured directionality of chiral coupling under different incident polarization handedness at two wavelengths ($\lambda_1$, $\lambda_2$) with opposite directionality. On the top of figure shows the incident polarization handedness that varies with rotation angle of QWP. The measured results in the experiment (balls) are in good agreement with predicated values by theory (dashed lines).

We characterize the directionality of chiral coupling in silicon photonic circuits based on Eq. (5) and measure power from two output ports of the Y-branch waveguide, as shown in Fig. 4. The adjustable polarization state of incident light by controlling the rotation angle of the QWP is illustrated on top of Fig. 4. The measured results under different incident polarization handedness at two wavelengths ($\lambda_1$, $\lambda_2$) with 3(g)-3(k) with opposite directionality. One can clearly see the distinct chiral coupling to different output ports of the Y-branch waveguide that is determined by the helicity of incident polarization handedness. Note that the measured intensity profiles from two output ports of the Y-opposite directionality are in good agreement with predicted values by theory. In particular, the absolute values of measured directionaliy $|D|$ exceed 0.92 under complete LCP ($\sigma^{-1}$) and RCP ($\sigma^{+1}$) inputs, indicating the achievable high directionality of chiral silicon photonic circuits.

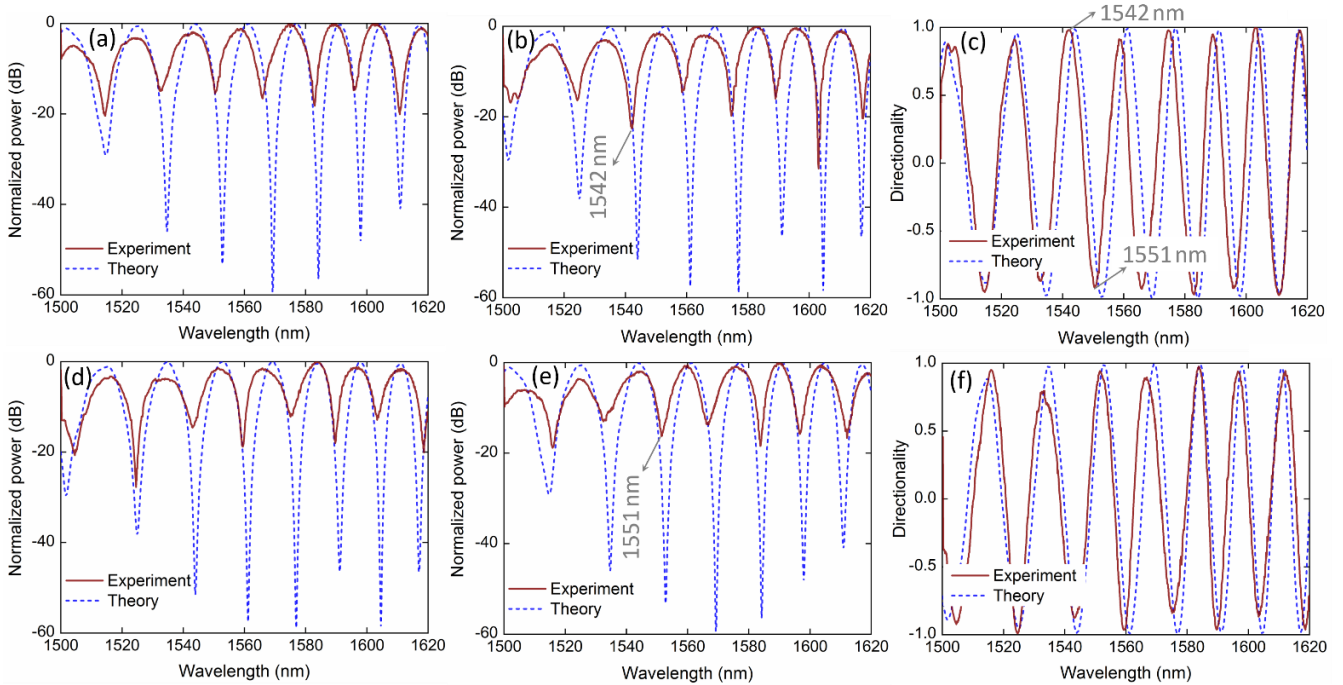

**Fig. 5.** Measured and simulated performance of chiral coupling in silicon photonic circuits versus wavelength. (a), (b), (d), (e) Normalized power from up (a, d) and down (b, e) output ports of the Y-branch waveguide. (c), (f) Directionality of chiral coupling in silicon photonic circuits. (a)-(c) LCP ($\sigma^{-1}$) incident light. (d)-(f), RCP ($\sigma^{+1}$) incident light. Solid lines: experiment. Dashed lines: theory.

We further study the performance of chiral coupling in silicon photonic circuits as a function of the wavelength. We measure the power from two output ports of the Y-branch waveguide and assess the directionality by sweeping the incident wavelength. For easy measurement of sweeping spectra, a pair of lensed fiber is used for fiber-chip-fiber coupling (see Supplement 1). The output lensed fiber is connected to an optical power meter to record the sweeping spectra. Figs. 5(a) and 5(b) show the measured normalized power from two output ports of the Y-branch waveguide, respectively, under incident LCP ($\sigma^{-1}$) light, while Figs. 5(d) and 5(e) plot the measured results under incident RCP ($\sigma^{+1}$) light. Figs. 5(c) and 5(f) depict the evaluated directionality of chiral coupling under incident LCP ($\sigma^{-1}$) and RCP ($\sigma^{+1}$) light, respectively. One can see the interesting quasi-periodic phenomena of sweeping spectra. This is because that the propagation constant difference ($\Delta\beta$) between $TE_0$ and $TE_1$ modes due to waveguide dispersion in Eq. (5) is wavelength dependent, leading to the variation of power and directionality with wavelength. Note that the selected two wavelengths (1542 nm, 1551 nm) with opposite directionality are marked in Figs. 5(b) and 5(e), respectively. To increase the wavelength range with high directionality, one may shorten the inversely tapered silicon waveguide (see Supplement 1). Moreover, it is possible to realize wavelength-tunable chiral coupling in practical applications assisted by thermal-optic tuning.

The simulation results in Figs. 5(c) and 5(f) show an ultra-high directionality approaching ±1, and the measured values of $|\sigma|$ can be over 0.98. For the efficiency of chiral coupling, the simulation results show that the insertion loss can be less than 30% (see Supplement 1). In the experiment, the measured insertion loss of the chiral silicon photonic circuits is estimated to be around -5 dB. The obtained results shown in Figs. 3-5 demonstrate the successful implementation of ultra-directional and high-efficient chiral coupling in silicon photonic circuits.

## 4. CONCLUSION

We propose and demonstrate simple silicon photonic circuits for on-chip chiral coupling with superior performance. The underlying mechanism of chiral coupling is low-order to high-order mode conversion and interference. The mode interference enables high directionality thanks to the complete destructive and constructive interference. The polymer-assisted coupling and guided-mode interference benefit high efficiency of chiral coupling with negligible scattering to non-guided modes. The presented high directionality and efficiency of chiral photonic behavior have not yet been achieved before based on other reported mechanisms. With future improvement, the directionality and efficiency could be further enhanced through the optimization of geometric parameters of the inversely tapered silicon waveguide structure and the bending radius of the Y-branch waveguide. Moreover, the chiral silicon photonic circuits can be potentially used as a perfect 3-dB power splitter for LP light with arbitrary polarization orientation, which has not yet been reported before in polarization-sensitive silicon nanophotonic devices. Apparently, the chiral silicon photonic circuits can be exploited to determinate the helicity of polarization handedness of light, opening a door to facilitate high-performance chip-scale photonic spin sorting and other spin-related applications. It is believed that photonic integrated circuits will play an increasingly important role in chip-scale chiral optics and more chirality-related emerging applications.

**Funding.** This work was supported by the National Natural Science Foundation of China (NSFC) under grants 61761130082, 11774116, 11574001, 11274131 and 61222502, the National Basic Research Program of China (973 Program) under grant 2014CB340004, the Royal Society-Newton Advanced Fellowship, the National Program for Support of Top-notch Young Professionals, the Natural Science Foundation of Hubei Province of China under grant 2018CFA048, and the Program for HUST Academic Frontier Youth Team.

See Supplement 1 for supporting content.